\documentstyle[preprint,aps]{revtex}

 \newcommand \be {\begin{equation}}
\newcommand \bea {\begin{eqnarray} \nonumber }
\newcommand \ee {\end{equation}}
\newcommand \eea {\end{eqnarray}}
 
 \newcommand \bi {\bibitem}
\newcommand \s {\sigma}
\newcommand \de {\delta}
\newcommand \De {\Delta}
\newcommand \g {\gamma}

\newcommand \la {\lambda}

 \newcommand \al {\alpha}

\begin{document}
\draft
\preprint{MA/UC3M/10/95}
\title{Analytical solution of the Monte Carlo dynamics of a simple
spin-glass model}

\author{L. L. Bonilla(*), F. G. Padilla(*), G. Parisi(**) 
and F. Ritort(*)}
\address{(*) Departamento de Matem\'aticas,\\
Universidad Carlos III, Butarque 15\\
Legan{\'e}s 28911, Madrid (Spain)\\
E-Mail: bonilla@ing.uc3m.es\\
E-Mail: padilla@dulcinea.uc3m.es\\
E-Mail: ritort@dulcinea.uc3m.es\
\\ (**) Dipartimento di Fisica,\\
Universit\`a di Roma I {\sl ``La  Sapienza''}
\\ INFN Sezione di Roma I \\ Piazzale
Aldo Moro, Roma 00187\\
E-Mail: parisi@vaxrom.roma1.infn.it}

\date{\today}
\maketitle

\begin{abstract}
In this note we present an exact solution of the Monte Carlo dynamics
of the spherical Sherrington-Kirkpatrick spin-glass model. We obtain
the dynamical equations for a generalized set of moments which can be
exactly closed. Only in a certain particular limit the dynamical
equation of the energy coincides with that of the Langevin dynamics
\end{abstract} 

\vfill
\pacs{64.70.Nr, 64.60.Cn}

\vfill

\narrowtext

There has been in recent years a renewal of the interest in the study
of the dynamics in spin glasses. The main motivation is based upon the
fact that real spin glasses (and also real glasses) are always off
equilibrium during the experimental time window, the most clear
signature being the existence of aging \cite{suecos}. Two main
approaches have been put forward very recently to understand this
problem. In the first approach, special emphasis is put on the
behavior of two-time quantities (like the correlation or the
response function at two different times) for a specific microscopic
dynamics\cite{CCC}. This has been complemented by the study of several
phenomenological models which try to capture the main essentials of
the slow dynamical process \cite{Bou}.  In the second approach, one
tries to find the time evolution of some macroscopic observables
(\mbox{one-time} quantities) and, eventually in a latter stage,
the evolution of the two-time quantities\cite{CS}. While less ambitious 
than the first approach, this line of thought allows one to obtain fairly
good results in simple cases.

The major part of these approaches have focused their attention in the
solution of the Langevin or Glauber dynamics\cite{Sommers}. In this
letter we analytically solve the Monte Carlo dynamics in a simple spin
glass model. There are two reasons why this study should be of interest.
First, there is no special reason to privilege a
particular type of dynamics over others, and it is important to understand 
why other dynamics may yield different results and how different these
results may be. The second reason is more practical and relies on the fact 
that the major part of numerical simulations use the Monte Carlo
algorithm. Consequently, more direct comparisons between theory and 
numerics can be done.

While the results we will show are based on a very simple spin-glass
model, it would be interesting to extend our approach to more complex
cases (for instance where replica symmetry is broken).
We will also  show that the Monte
Carlo dynamics is different from the Langevin dynamics although the  same
dynamical equation for the energy is obtained in a certain limit.

{\em The model and the dynamics}. The model we are considering is the
two-spin spherical spin-glass model \cite{KTJ}, defined by
\be
E\lbrace\s\rbrace=-\sum_{i<j}\,J_{ij}\s_i\s_j
\label{eq1}
\ee

\noindent
where the indices $i,j$ run from $1$ to $N$ ($N$ is the number of
lattice sites) and the spins $\s_i$ satisfy the spherical global 
constraint
\be
\sum_{i=1}^N\,\s_i^2=N.
\label{eq2}
\ee

The interactions $J_{ij}$ are Gaussian distributed with zero mean and
$1/N$ variance. This model has been extensively studied in the
literature in all its details (the statics and the Langevin dynamics
\cite{MCC}) and is an useful starting point for our approach.

We will consider the Monte Carlo dynamics with the Metropolis algorithm
(another algorithm would yield the same qualitative results). 
The dynamics is done in this way: we take the configuration
$\lbrace\s_i\rbrace$ at time $t$ and we perform a small random rotation of
that configuration to a new configuration $\lbrace\tau_i\rbrace$ where
\be
\tau_i=\s_i+\frac{r_i}{\sqrt{N}}
\label{eq3}
\ee
and the $r_i$ are random numbers
extracted from a Gaussian distribution $p(r)$ of finite 
\mbox{variance $\delta$,}
\be
p(r)=\frac{1}{\sqrt{2\pi\delta^2}}\exp(-\frac{r^2}{2\delta^2})~~~.
\label{eq4}
\ee

We impose the new configuration $\lbrace\tau_i\rbrace$ to satisfy the
spherical constraint eq.(\ref{eq2}). Let us denote by $\Delta E$ the
change of energy $\Delta E=E\lbrace\tau\rbrace-E\lbrace\s\rbrace$.
According to the Metropolis
algorithm we accept the new configuration with probability 1 if 
$\Delta E< 0$
and with probability $exp(-\beta\Delta E)$ if $\Delta E> 0$
where $\beta=\frac{1}{T}$ is the inverse of the temperature $T$.

We have chosen the particular equation of motion (\ref{eq3}) 
because it makes the dynamics invariant under rotations. There are other
types of motions, for instance moving only one randomly chosen component
$\tau_i=\s_i+r_i$, but they do complicate much more the analytical
treatment (details will be shown elsewhere \cite{nostre}). 

{\em The joint probability} $P(\Delta h_k,\Delta E)$.  Because
the dynamics eq.(\ref{eq3}) is invariant under rotations we will work in
what follows in the diagonal basis of the interaction matrix
$J_{ij}$. In that basis the energy reads,

\be
E\lbrace\s_{\la}\rbrace=-\sum_{\la}J_{\la}\s_{\la}^2
\label{eq5}
\ee 
where the $\s_{\la}$ are the eigenvectors and the $J_{\la}$ are
distributed according to the Wigner semicircular law \cite{WIGNER}, \be
w(\la)=\frac{\sqrt{4-\la^2}}{2\pi}.
\label{eq6}
\ee

We also define the generalized $k$-moments,
\be
h_k=\sum_{(i,j)}\s_i(J^k)_{ij}\s_j=\sum_{\la}J_{\la}^k\s_{\la}^2
\label{eq7}
\ee

\noindent
where $h_0=1$ (spherical constraint) and $h_1=-2E$.  The basic object we
want to compute is the joint probability $P(\Delta h_k,\Delta E)$
to have a certain variation $\De h_k$ of the $k$-moment given that the
energy $E$ has also varied by a quantity $\De E$. This is a 
quantity  which gives all the information about the dynamics.
The variation of the quantities $h_k$ and $E$ in an elementary move
eq.(\ref{eq3}) are given by
\bea
\De
E^*=-\frac{1}{\sqrt{N}}\sum_{\la}J_{\la}\s_{\la}r_{\la}-
\frac{1}{2N}\sum_{\la}\,J_{\la}\,r_{\la}^2\\
\De h_k^*=\frac{2}{\sqrt{N}}\sum_{\la}J_{\la}^k\,\s_{\la}\,r_{\la}+
\frac{1}{N}\sum_{\la}\,J_{\la}^k\,r_{\la}^2.
\label{eq8}
\eea

The joint probability $P(\Delta h_k,\Delta E)$ is,
\be
P(\Delta h_k,\Delta E)=\int\de(\De h_k-\De h_k^*) \de(\De E-\De E^*)
\de(\De h_0)\prod_{\la} \Bigr( p(r_{\la})dr_{\la}\Bigl)
\label{eq9}
\ee
where the last delta function in the integrand accounts for the
spherical constraint and the variations $\De h_k^*,\De E^*$ are given in
eq.(\ref{eq8}). 

Using the integral representation for the delta function 
\be
\de(x)=\frac{1}{2\pi}\int_{-\infty}^{\infty}e^{i\alpha x}d\alpha
\label{eq10}
\ee

\noindent
and substituting in (\ref{eq8}) we get
\bea
P(\Delta h_k,\Delta E)=\int d\alpha\, d\mu\, d\eta\, 
\exp\Bigl (i\alpha\De h_k+
i\mu\De E -\\
\frac{\de^2}{2N}\sum_{\la}\frac{\s_{\la}^2\g_{\la}^2}
{(1-\frac{i\g_{\la}^2
\de^2}{N})}-\frac{1}{2}\sum_{\la}\log(1-\frac{i\g_{\la}^2
\de^2}{N})\Bigr )
\label{eq11}
\eea
\noindent

\noindent
where $\g_{\la}=-2\al J_{\la}^k+\mu J_{\la}+2\eta$. Expanding the
logarithm and retaining the first $1/N$ correction we get (after some
 manipulations)
\be
P(\Delta h_k,\Delta E)=P(\De E)\,P(\Delta h_k|\Delta E)
\label{eq12}
\ee
where $P(\De E)$ is the probability distribution to have a change of
energy $\De E$ and $P(\De h_k|\De E)$ is the conditional probability
of $\De h_k$ given $\De E$. The final expressions are,

\bea
P(\De E)=\frac{1}{\sqrt{2\pi\de^2 B_1}}\,\exp\Bigl (-\frac{(\De E+\de^2 E)^2}
{2\de^2B_1}\Bigr )\\
P(\Delta h_k|\Delta E)=\frac{1}{\sqrt{8\pi(C_k-(B_k^2/B_1))}}\,
\exp(-\frac{\De h_k+\de^2(h_k-<<J^k>>)+2\frac{B_k}{B_1}(\De
E+\de^2 E)^2}{8\de^2(C_k-B_k^2/B_1)}\Bigr )
\label{eq13}
\eea
\noindent
with
\bea
C_k=h_{2k}-h_{k}^2;~~~B_k=h_{k+1}+2Eh_{k};~~~(h_0=1;h_1=-2E);~~~
\\<<f(J)>>=\int_{-2}^{2}d\la w(\la)f(\la);
\label{eq14}
\eea

Before showing the dynamical equations for the moments we will prove
that equilibrium is a stationary solution of the Monte Carlo dynamics.
The equation for the energy is obtained by considering the average
variation of energy in an elementary move,
\be
\overline{\De E}=\int_{-\infty}^{0}\De E\,P(\De E)d\De E+
\int_{0}^{\infty}\De E\,\exp(-\beta \De E)P(\De E)d\De E~~~~~.
\label{eq15}
\ee

A direct calculation shows that this variation is zero when $B_1=
h_2-4E^2=-2ET$. It can be easily shown (using standard static
calculations \cite{MePaVi}) that this is the condition satisfied at
equilibrium.

Also one can compute the acceptance rate as a function of time, which is
given by
\be
A(t)=\int_{-\infty}^{0}P(\De E)d\De E+
\int_{0}^{\infty}\exp(-\beta \De E)P(\De E)d\De E~~~~~.
\label{eq16}
\ee

In what follows we will consider the zero temperature case, the computations
being straightforward in case of finite temperature.
A straightforward computation shows,
\be
A(t)=\frac{Erf(\alpha)}{2}
\label{eq17}
\ee

\noindent
where $Erf(\alpha)$ is the error function
$Erf(\alpha)=\frac{2}{\sqrt{\pi}}
\int_{\alpha}^{\infty}\,dx\exp(-x^2)$ and the parameter $\alpha$ is
given by,
\be
\alpha=-\frac{\de E}{\sqrt{2B_1}}
\label{eq18}
\ee

Now we can understand qualitatively how the dynamics goes on. Suppose we
start at zero temperature with a random initial configuration $\s_i=\pm
1$ such that $E(t=0)=0$ and $B_1(t=0)=1$. The energy monotonically
decreases to the ground state energy $E=-\frac{J_{max}}{2}=-1$ while
$B_1$ decreases also to zero. In the large time limit $\alpha$ diverges
and the acceptance rate goes to zero (we are at zero temperature).
There are two different regimes in the dynamics. The first one is an initial
regime where $\alpha$ is small and the acceptance rate is nearly
$1/2$. This corresponds to a gaussian $P(\De E)$ (eq.(\ref{eq13})) with
width $\delta \sqrt{B_1}$  larger than the position of its center
($\delta^2E$). In this case, the changes of configuration which
increase or decrease the energy have the same probability. The energy 
decreases fast in this regime because the acceptance is large.
 The second regime appears when $B_1$ is so small in order that
$\alpha$ becomes large. In this case the acceptance is very small (it
goes like $\frac{exp(-\alpha^2)}{\al}$) and the dynamics is strongly
slowed down. The system goes very slowly to the equililibrium.

In order to obtain the time evolution of the acceptance rate we need to
know the energy $E$ and $B_1$ at time $t$. In the next paragraph we will
show that all $k$-moments only depend on these two quantities.  In
figure 1 we show the results for the acceptance rate obtained in a
self-consistent way using expression (\ref{eq17}). The two regimes
(separated by a drastic fall of the acceptance rate $A(t)$) can be clearly
appreciated.

{\em Analytical solution of the hierarchy}. In order to obtain the
dynamical evolution of the $k$-moments $h_k$ we have to compute its
average variation in a Monte Carlo step over the accepted changes of
configuration. In this case one Monte Carlo step corresponds to $N$
elementary moves. In the thermodynamic limit we can write the continuous
equations,

\newpage
\bea
\frac{\partial h_k}{\partial t}=\overline{\De h_k}=
\int_{-\infty}^{\infty}\De h_k\,d\De h_k\,\\ \Bigl (\int_{-\infty}^{0}d\De E 
 P(\De h_k,\De E)+\int_{0}^{\infty}d\De E
\,\exp(-\beta\De E) P(\De h_k,\De E)\Bigr )
\label{eq19}
\eea

In the zero-temperature case one obtains,

\be
\frac{\partial h_k}{\partial t}=-\frac{\de^2(h_k-<<J^k>>)
Erf(\alpha)}{2}-\frac{2}{\sqrt{\pi}}\alpha B_k E \exp(-\alpha^2)
\label{eq20}
\ee

\noindent
where the average $<<..>>$ has been previously defined in eq.(\ref{eq14}).
In particular one gets, for $k=0$, 
$\frac{\partial h_0}{\partial t}=0$ which is the spherical
constraint. For the energy $E=-\frac{h_1}{2}$ we get the equation,

\be
\frac{\partial E}{\partial t}=\frac{B_1}{E} K(\alpha)
\label{eq21}
\ee

\noindent
where $K(\alpha)=\frac{\al\exp(-\al^2)}{\sqrt{\pi}}-
\al^2\,Erf(\al)$ and $B_{1}=h_{2}^2-4 E^2$, where $h_{2}$ is the 
squared local field. In the first dynamical regime ($\alpha$ small) we get
$\frac{\partial E}{\partial t}=-\frac{\de\sqrt{B_1}}{\sqrt{2\pi}}$ and
in the slow dynamical regime ($\alpha$ large) we find $\frac{\partial
E}{\partial t}=\frac{B_1 exp(-\alpha^2)} {2E\alpha \sqrt{\pi}}$. In the
last case, if we redefine the time $\tau=tA(t)$ then we obtain the
expression $\frac{\partial E}{\partial \tau}=\frac{B_1}{E}=-B_1$
(because $E=-1$ for large enough times).  In this limit we get the
equation for the energy in the Langevin dynamics (\cite{nostre}).

\be
g(x,t)=\sum_{(i,j)}\s_i\,(e^{xJ})_{ij}\,\s_j=\sum_{\la} e^{\la x}\s_{\la}^2(t).
\label{eq22}
\ee

\noindent
This function yields all the moments $h_k=\Bigl (\frac{\partial^k
g(x,t)}{\partial x^k}\Bigr )_{x=0}$.

It is easy to check that $g(x,t)$ satisfies the following differential
equation

\be
\frac{\partial g(x,t)}{\partial t} =a(t)\frac{\partial g(x,t)}{\partial
x}+b(t)g(x,t)+c(x,t)
\label{eq23}
\ee

where the coefficients are given by,
\be
a(t)=-\frac{2\,\al\,e^{-\al^2}}{E\sqrt{\pi}}\\
\label{eq222}
\ee
\be
b(t)=-(\frac{\de^2\,Erf(\alpha)}{2}+\frac{4\al\,\exp(-\al^2)}{\sqrt{\pi}})\\
\label{eq232}
\ee
\be
c(x,t)=\frac{\de^2\,<<e^{xJ}>>\,Erf(\al)}{2}
\label{eq24}
\ee

\noindent
where $a(t)$ is a positive quantity.
Note the difference with the Langevin case in which
$a(t)=2,\,b(t)=4E,\,c(x,t)=0$ \cite{nostre}.  
The solution of this partial differential equation
with the initial conditions $g(0,t)=1$,
$g(x,0)=<<\exp(x\la)\,\s^2(\la,t=0)>>$ and subject to the self-consistency
conditions (\ref{eq222}-\ref{eq24}) where 
$\al$ is given by

\be
\al=\frac{\de \frac{\partial g}{\partial x}(0,t)}{2\sqrt{
\frac{\partial^2 g}{\partial x^2}(0,t)-
\Bigl (\frac{\partial g}{\partial x}(0,t)}\Bigr )^2}
\label{luis}
\ee

\noindent
is

\bea g(x,t)=<<\exp^{(x+\int_0^t
a(t')dt')\la}\s^2(\la,t=0)>>\, \exp(\int_0^t b(t')dt')+\\
\delta^2\int_0^t\,dt' c(x+\int_{t'}^t a(t'')dt'',t')\exp(\int_{t'}^t
b(t'')dt'').
\label{eq25}
\eea

>From this function we can
readily obtain all moments as a function of time. In particular, we show
in figure 2 the energy obtained (at zero temperature) in a real Monte
Carlo simulation as a function of time compared to the theoretical
prediction obtained from the previous equation. 

We note the following differences between Monte Carlo and Langevin
dynamics.  In the Langevin dynamics one can show that the time evolution
of all $k$-moments is completely determined only by
the energy (the first moment). In the Monte Carlo case we have seen 
the time evolution of the moments
is determined by the time dependent parameter $\alpha$ which is a
function of the energy $E$ and the second cumulant as shown in eq. (\ref{eq25}).
 In this sense the dynamics is slightly more complicated than the Langevin case
but simple enough to be governed by two (time dependent) quantities.

Now we can summarize our results. We have analytically solved the
Monte Carlo dynamics of a simple spin-glass model (without replica
symmetry breaking). The method consists in constructing the joint
probability eq.(\ref{eq9}) of having a certain change of the
generalized moments $h_k$ for a given change $\De E$ of energy. Once
this probability is constructed it is possible to derive the dynamical
evolution equations for all moments. The hierarchy of equations can be
closed by introducing the generating functional $g(x,t)$. While we
have applied this method in a very simple case we expect it to be
aplicable to other more interesting cases where replica symmetry
is broken. The philosophy of the method is very close to that devised
by Coolen and Sherrington\cite{CS} but applied in our case for the the
specific Monte Carlo dynamics. It is interesting to note that the
hierarchy of equations (\ref{eq20}) is different from that obtained
in case of Langevin dynamics. Only in the large time limit, and
renormalizing the time by the acceptance ratio, the equation for the
energy coincides in both cases (but higher moments do not
coincide). It would be also interesting to try to derive the
correlation functions and the response function in this framework.

We are indebted to Silvio Franz for useful discussions on these
subjects. (F.R.) and (L.B.) acknowledge Ministerio de Educaci\'on
y Ciencia and European Comunity for finacial support through grant
PB92-0248.

\vfill\eject

\vfill
\newpage
{\bf Figure Captions}
\begin{itemize}

\item[Fig.~1] Acceptance rate $A(t)$ calculated self-consistently 
using eq.(\ref{eq17}) compared to Monte Carlo results at zero temperature 
for two different values of $\delta=0.1$ (rhombs) and $0.01$ (crosses) 
for $N=2000$ and $N=500$ respectively.

\item[Fig.~2] Relaxation of the internal energy as a function of time
for three different values of $\delta$ ($0.1$ (rhombs),$0.01$(crosses),
 $0.001$ (boxes)) and $N=2000$ at zero temperature compared with the
analytic prediction eq.(\ref{eq25}). 

\end{itemize}

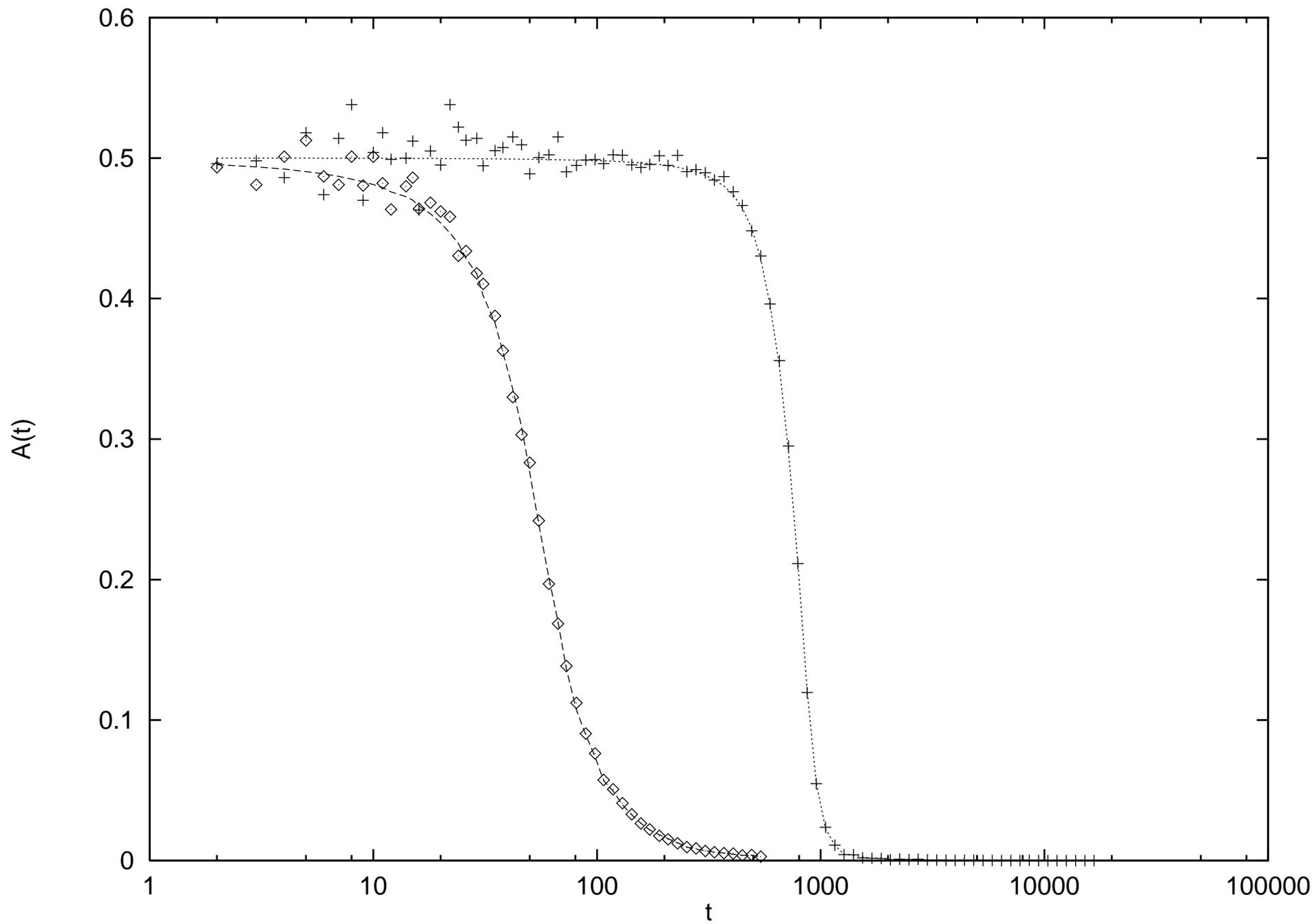
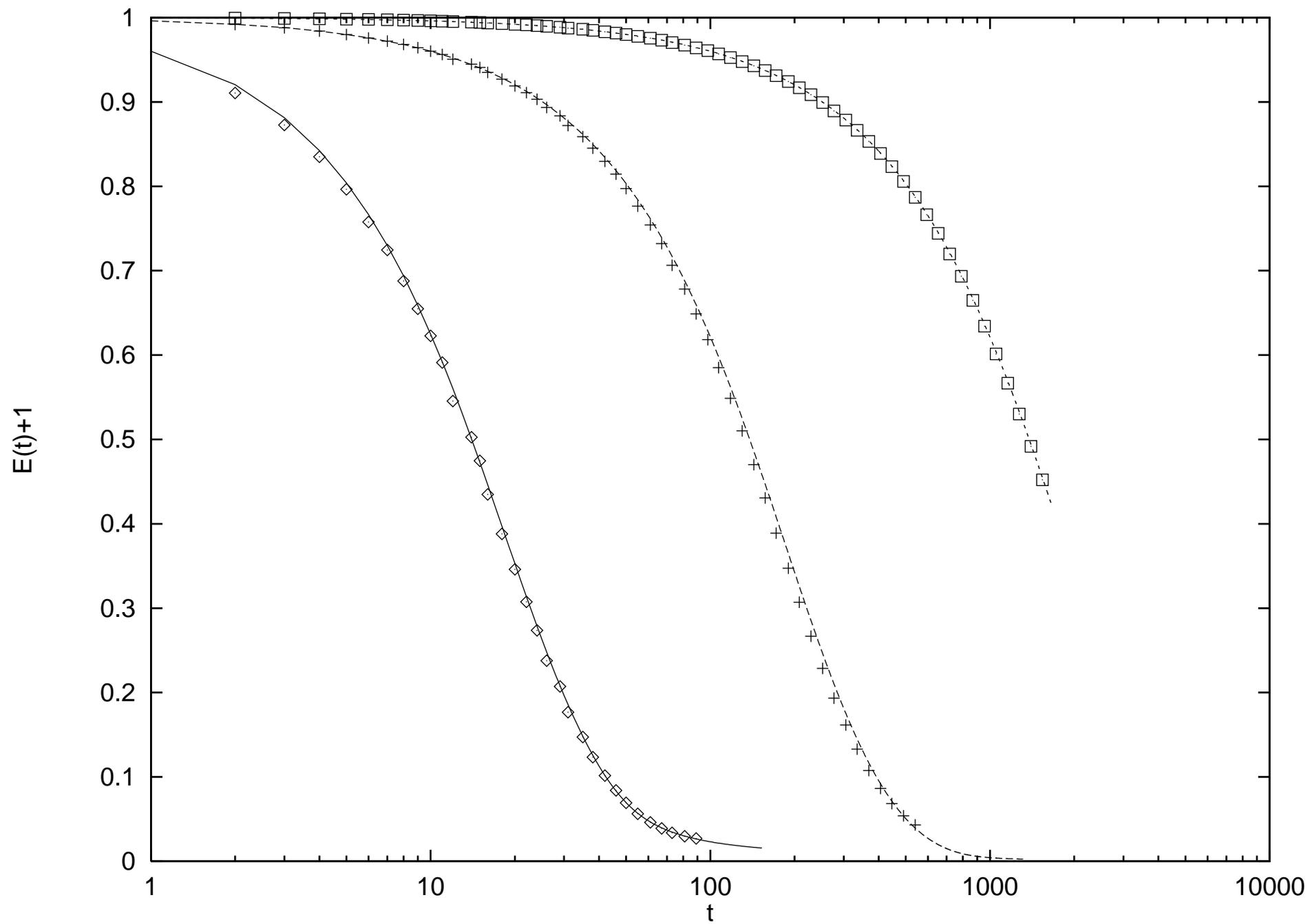
\end{document}